\documentclass[twocolumn,twocolappendix]{aastex631}

\shorttitle{Graduate Student Input to Faculty Hiring}
\shortauthors{Yale ASC et al.}
\graphicspath{{./}{figures/}}

\usepackage[normalem]{ulem}
\usepackage{enumitem}
\usepackage{bm}
\usepackage{amsmath}
\usepackage{gensymb}
\usepackage{subfigure}
\usepackage{multirow}
\usepackage{array}
\usepackage{hhline}
\PassOptionsToPackage{hyphens}{url}\usepackage{hyperref}

\begin{document}

\title{A Standardized Framework for Collecting Graduate Student Input in Faculty Searches}


\author[0000-0002-8320-2198]{Yasmeen Asali}
\altaffiliation{}
\affiliation{Department of Astronomy, Yale University, 52 Hillhouse Ave, New Haven, CT 06511, USA}
\affiliation{Astronomy Student Council (ASC) at Yale, 2021/22}

\author[0000-0002-4836-1310]{Konstantin Gerbig}
\altaffiliation{}
\affiliation{Department of Astronomy, Yale University, 52 Hillhouse Ave, New Haven, CT 06511, USA}
\affiliation{Astronomy Student Council (ASC) at Yale, 2021/22}

\author[0000-0002-2525-9647]{Aritra Ghosh}
\altaffiliation{}
\affiliation{Department of Astronomy, Yale University, 52 Hillhouse Ave, New Haven, CT 06511, USA}
\affiliation{Astronomy Student Council (ASC) at Yale, 2021/22}

\author[0000-0001-8722-1436]{Christopher Lindsay}
\altaffiliation{}
\affiliation{Department of Astronomy, Yale University, 52 Hillhouse Ave, New Haven, CT 06511, USA}
\affiliation{Astronomy Student Council (ASC) at Yale, 2021/22}

\author[0000-0002-5120-1684]{Zili Shen}
\altaffiliation{}
\affiliation{Department of Astronomy, Yale University, 52 Hillhouse Ave, New Haven, CT 06511, USA}
\affiliation{Astronomy Student Council (ASC) at Yale, 2021/22}

\author[0000-0002-7007-9725]{Marla Geha}
\altaffiliation{}
\affiliation{Department of Astronomy, Yale University, 52 Hillhouse Ave, New Haven, CT 06511, USA}

\collaboration{6}{( \textsuperscript{*}Equal Contribution )}

\begin{abstract}

We present a procedure designed to standardize input received during faculty searches with the goal of amplifying student voices.  The framework was originally used to collect feedback from graduate students, but it can be adapted easily to collect feedback from undergraduate students, faculty, staff or other stakeholders.   Implementing this framework requires agreement across participating parties and minimal organization prior to the start of faculty candidate visits.

\end{abstract}


\section{Introduction} \label{sec:intro}

Faculty hiring at major research universities includes input from many different groups. Graduate students are highly impacted by available faculty in a department and can provide critical and unique feedback during the hiring process. In a living document written by members of Colorado State University's Graduate Center for Inclusive Mentoring and Graduate Student Council titled \cite{CSU_doc}, the authors also argued that high levels of graduate student involvement can reflect well on the department itself, thus potentially attracting higher quality candidates. Further, graduate students themselves benefit from the professional development experience that is active participation in a hiring process. For these reasons, graduate student input is considered best practice in faculty hiring processes \citep[see e.g.,][]{Columbia, Cornell}. 

However, the degree of graduate student involvement varies significantly between university departments, and can encompass for example student representation on the hiring committee \citep{Rutgers} or the collection of graduate input from web-based feedback forms \citep{Washington}.

Inspired by a survey designed during a 2022 faculty hiring process at Yale University, we present a procedure to amplify graduate student voices in the hiring process. This write-up and associated survey documentation is intended to be easily adopted by other departments to standardize and improve the faculty hiring process.

In 2022, the Department of Astronomy at Yale University initiated a search for a new faculty member. Current graduate students were given the opportunity to interact with all the shortlisted candidates via an hour-long meeting, as well as the candidate's colloquium and other informal departmental interactions.   To reduce subjectivity, the Yale Astronomy Student Council (ASC) designed and implemented a procedure to gather and present data collected from all graduate students during these interactions. The ASC is an elected body of 3 to 5 graduate student members that work to address a wide variety of issues concerning graduate students and the graudate program within the department. These data and an associated write-up were presented at a department faculty meeting at the end of the candidate interview process. 
 
This paper was inspired by the positive response to the faculty hiring survey from both students and faculty.   Although we cannot quantify its impact, the survey provided a more standardized method to provide student feedback into the decision making process compared to previous faculty hires at Yale Astronomy.  It provided a clear picture of graduate student opinions, highlighting both points of consensus and areas of disagreement.

With the goal of reducing the barrier for others to adopt these more objective evaluation methods, we provide an overview of the survey design, implementation and data presentation.  Example survey forms and python code to analyze results are provided via a public GitHub repository (see Appendix \ref{ap:sec:data_release}).

We note that this entire work was carried out at an American R1 research university with its own specific history, bias, and socio-economic existence. It might be important to take this into consideration while adapting/implementing this procedure at a different institution.

\section{Survey design and implementation}\label{sec_design}

Our framework for graduate student input in faculty hiring was developed and is intended for a procedure that includes multi-day visits for each short-listed candidate.   The process must be set in motion prior to the first candidate's arrival.   The first step must be to establish a consensus about the structure of the visits, such that they are run homogeneously across faculty candidates.

\begin{figure*}[htb]
    \centering
    \includegraphics[width = \textwidth]{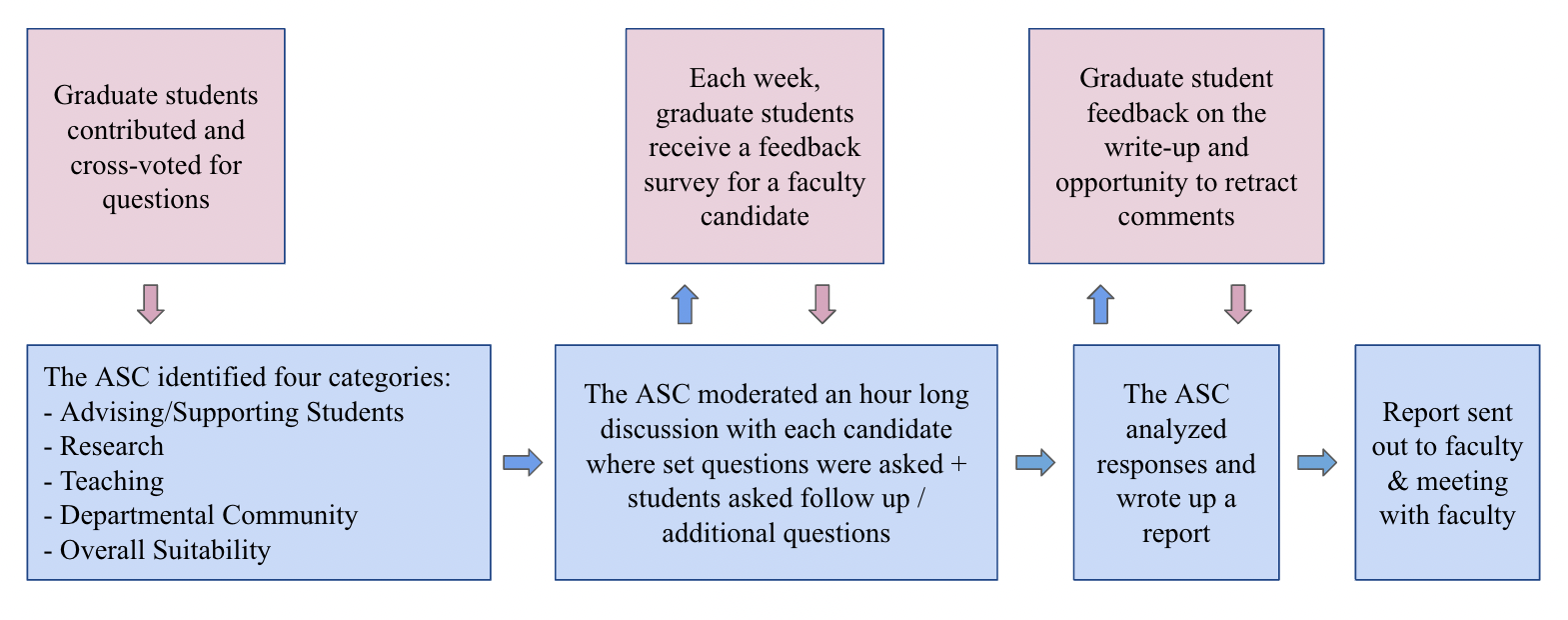}
    \caption{Flowchart showing our procedure for collecting feedback on the faculty candidates from the graduate students. Blue boxes indicate ASC actions and red boxes indicate input from the full graduate student body.}
    \label{fig:procedure}
\end{figure*}

In our case, the graduate student body was invited to department-wide science colloquia presented by the faculty candidates. In addition, there were also scheduled, one hour long zoom meetings open exclusively to the candidate and graduate students, which were moderated by ASC members. These two events constituted the basis for graduate students to form an opinion and provide feedback on the candidates. Two weeks after the last candidate's interview, there was a meeting with the graduate student body and current faculty, where the results of the survey were presented. The entire process was conducted fully virtually in light of the Covid19 pandemic. A summary flowchart of our method for collecting graduate student feedback is shown in Fig.~\ref{fig:procedure}.

\subsection{Selecting interview questions}\label{ssec_questions}

During each meeting, the ASC asked the candidate a number of set questions, with time allocated for both candidate response and potential spontaneous follow-up questions from graduate students. The set list of questions was determined prior to the first candidate's interview and remained unchanged for the duration of the hiring process. Graduate students were first asked to contribute and then to cross-vote for questions in order to establish a ranking. From the list of questions, the ASC identified four general areas of interest: advising students, teaching (specifically graduate classes), departmental community, and research. Appendix~\ref{sect:list_of_questions} contains the developed list of questions.

Note that, in order to avoid overlap with the candidates' colloquia, where they discussed their respective research programs and scientific goals, the graduate student questions were intended to focus more on the candidates' thoughts on advising, teaching, developing departmental community, and their specific impact on the student body. 

\subsection{Structure of the graduate student - faculty candidate meeting}

The graduate student - faculty candidate meetings started with a brief introduction by the faculty candidate. The remaining time was divided into four sections corresponding to the previously identified areas of interests. The ASC always asked the faculty candidate the question with highest number of votes first before moving on to the questions with the next highest numbers of votes until time ran out on that area of interest. This ensured that candidates, while being confronted with varying numbers of questions, all answered the same 2-3 top voted questions within each area of interest. 

Finally, the last 10 minutes were specifically allocated for the candidate to ask the graduate student body questions, which we found to be particularly useful for graduate students to gauge the faculty candidate's priorities.

\subsection{Data collection}
\label{sect:data_collection}

The ASC sent out a link to a Google form survey after each candidate had given their colloquium. Graduate students were required to log in with their Yale email addresses to verify their identities. However, their emails were not recorded with their responses in order to preserve anonymity. Graduate students could indicate whether they were above or below Year 2 at the time of the survey, corresponding to whether they were taking classes or had advanced to candidacy, although we did not end up using this information in the final analysis.

Students graded each candidate on a 0-5 scale (with 0 being ``absolutely don't hire'' and 5 being ``strong hire'') for each of the four identified areas of interest. Finally, each graduate student was asked to provide an overall assessment, again from 0-5. Students also had the opportunity to write long-form responses if they had additional comments about the candidate for each section. 

We encouraged the graduate students to fill out the survey for each candidate before the subsequent week's faculty candidate - graduate student meeting but kept each survey form open in case students would like to edit their responses, and to increase participation. We note that keeping the survey form open after the week of each candidate's visit introduces the problem of recency bias and leads to graduate students comparing two faculty candidates directly if filling out two survey forms at the same time. 

All feedback forms were closed one week after the last candidate's interview. Leading up to this deadline, the ASC sent out reminder emails to the graduate student body to fill out the survey forms. To encourage graduate student participation and mitigate low number statistics, these emails included the current number of responses for each candidate.

\section{Survey analysis and mock results}\label{sec_results}

\begin{figure*}[htb]
    \centering
    \includegraphics[width = 0.75\textwidth]{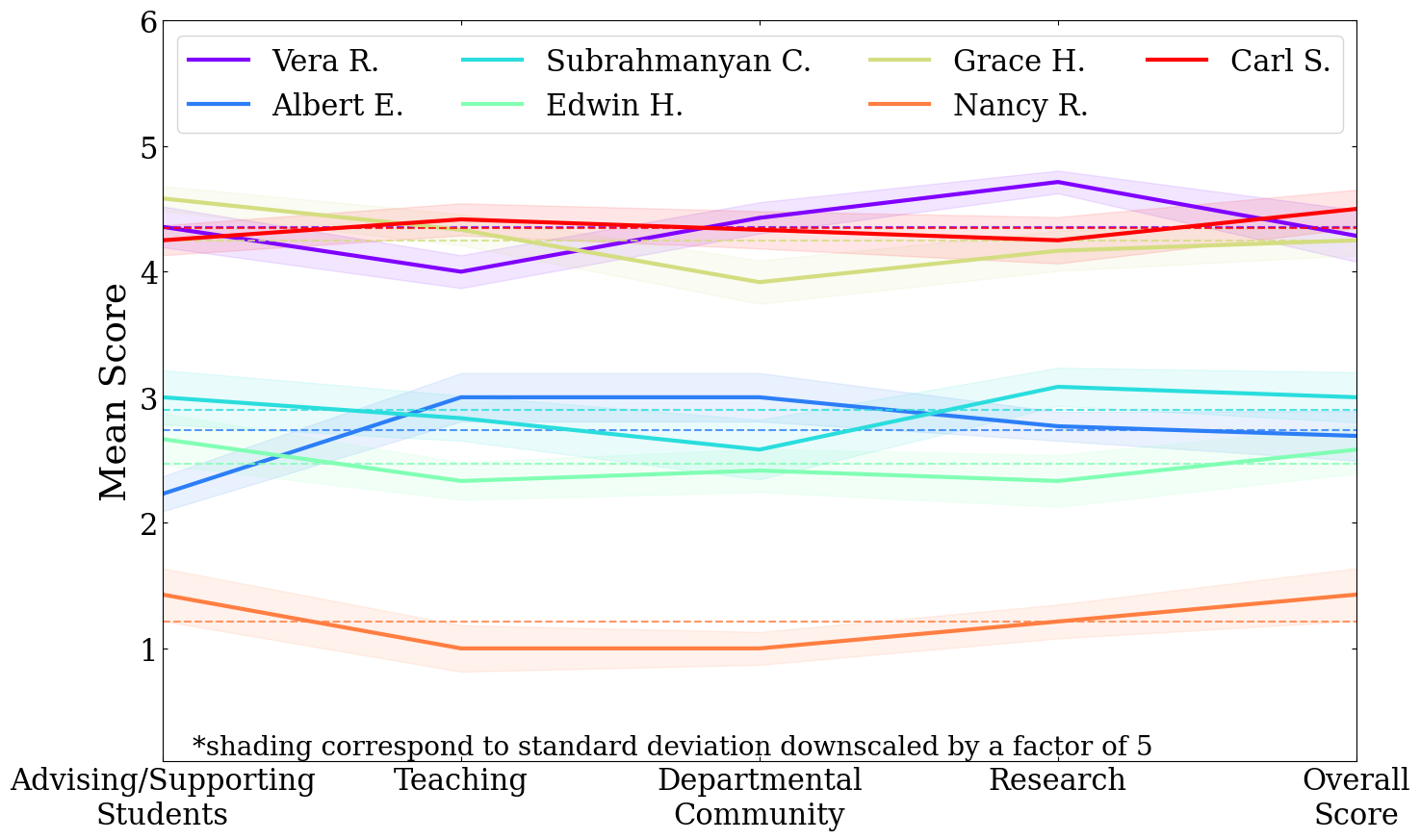}
    \caption{The mean values of the grades assigned by graduate students on different parameters for all the candidates. The dashed lines represent the average values of scores for individual candidates. The shown results are based on randomly generated mock data. }
    \label{fig:mean_all_scores}
\end{figure*}

\begin{figure*}[htb]
    \centering
    \includegraphics[width = 0.75\textwidth]{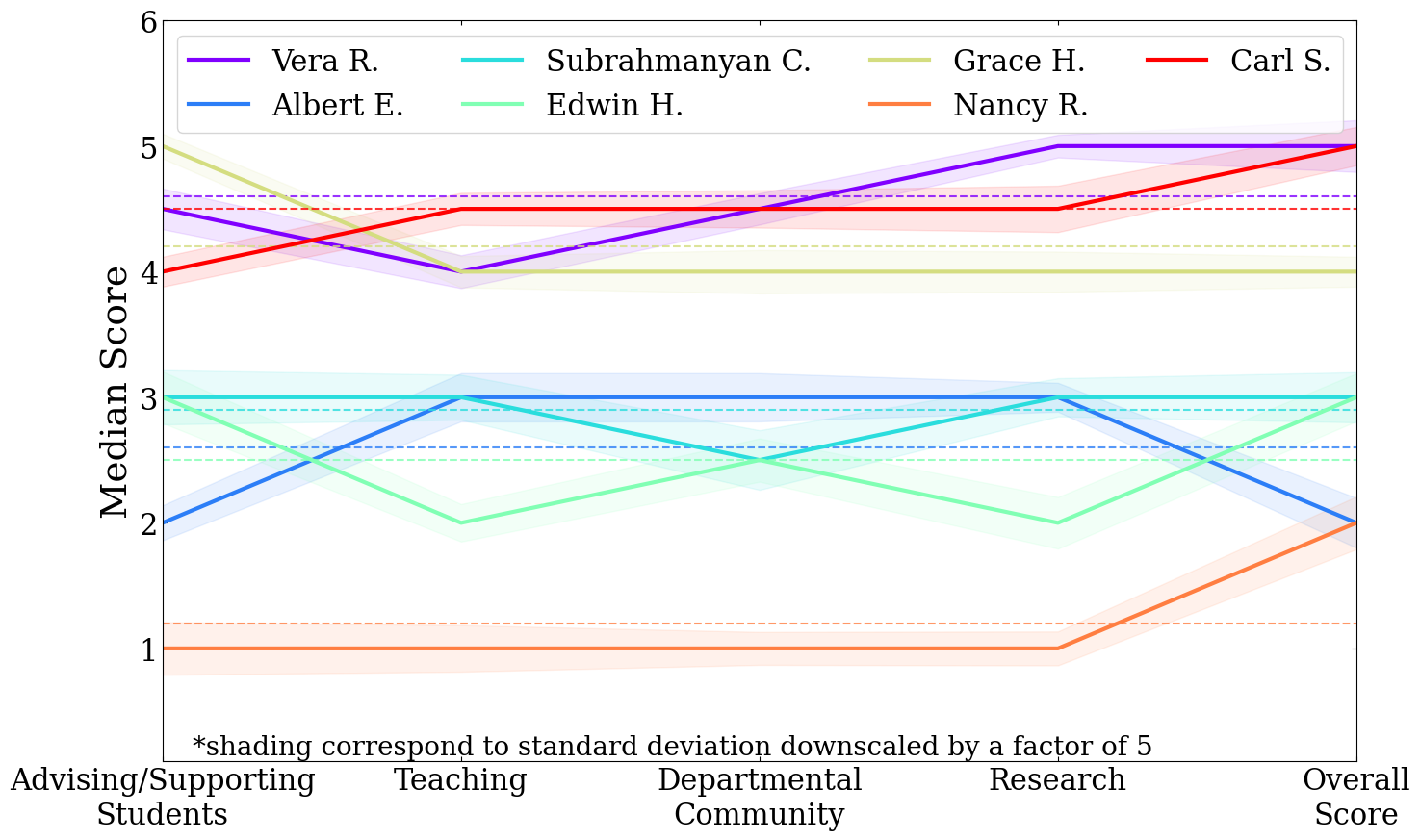}
    \caption{Same as Figure \ref{fig:mean_all_scores}, but now showing the median values of the scores instead of the mean. The shown results are based on randomly generated mock data.}
    \label{fig:median_all_scores}
\end{figure*}

\begin{figure*}[htb]
    \centering
    \includegraphics[width = 0.85\textwidth]{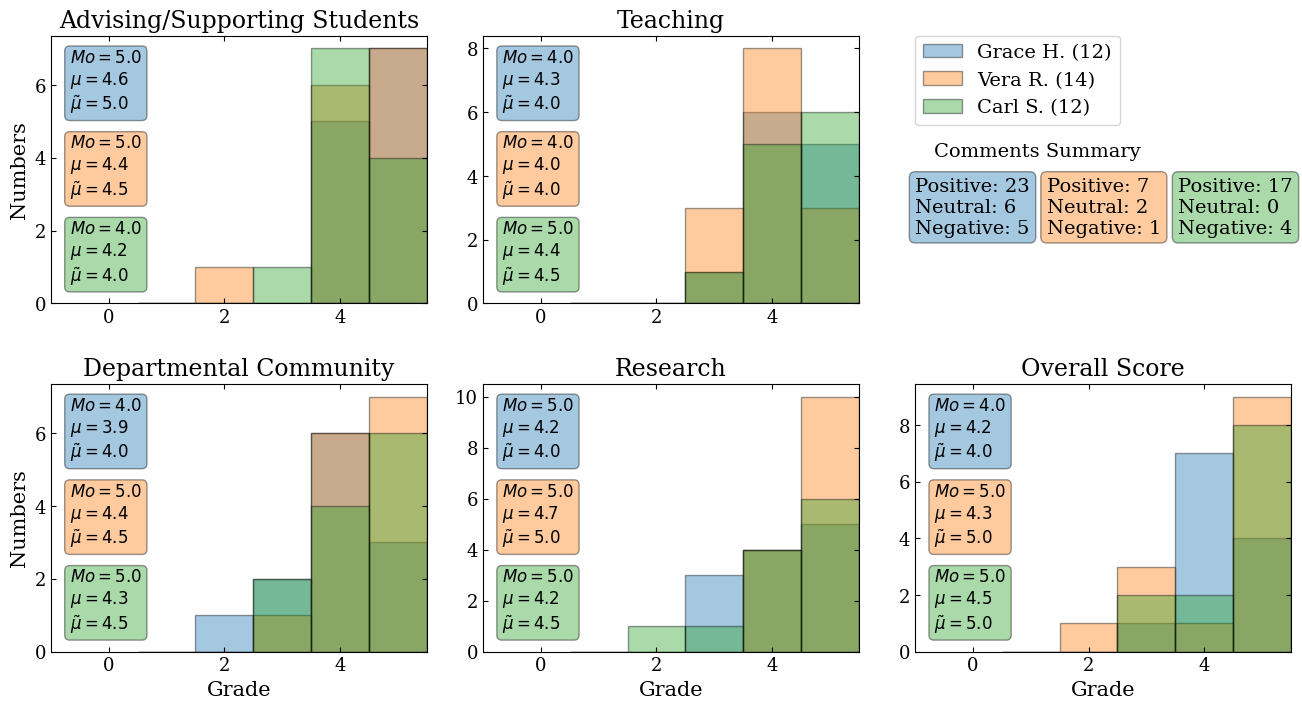}
    \caption{Histograms of grades for Grace H.,Vera R., and Carl S. The numbers in the legend refer to the number of respondents for each candidate. The mean($\mu$), median($\tilde{\mu}$), and mode (Mo) of each distribution are shown in the top-left of each panel. The number of comments for each respondent, separated into positive, neutral, and negative, is shown on the right.  The shown results are based on randomly generated mock data.}
    \label{fig:hist_1}
\end{figure*}

\begin{figure*}[htb]
    \centering
    \includegraphics[width = 0.85\textwidth]{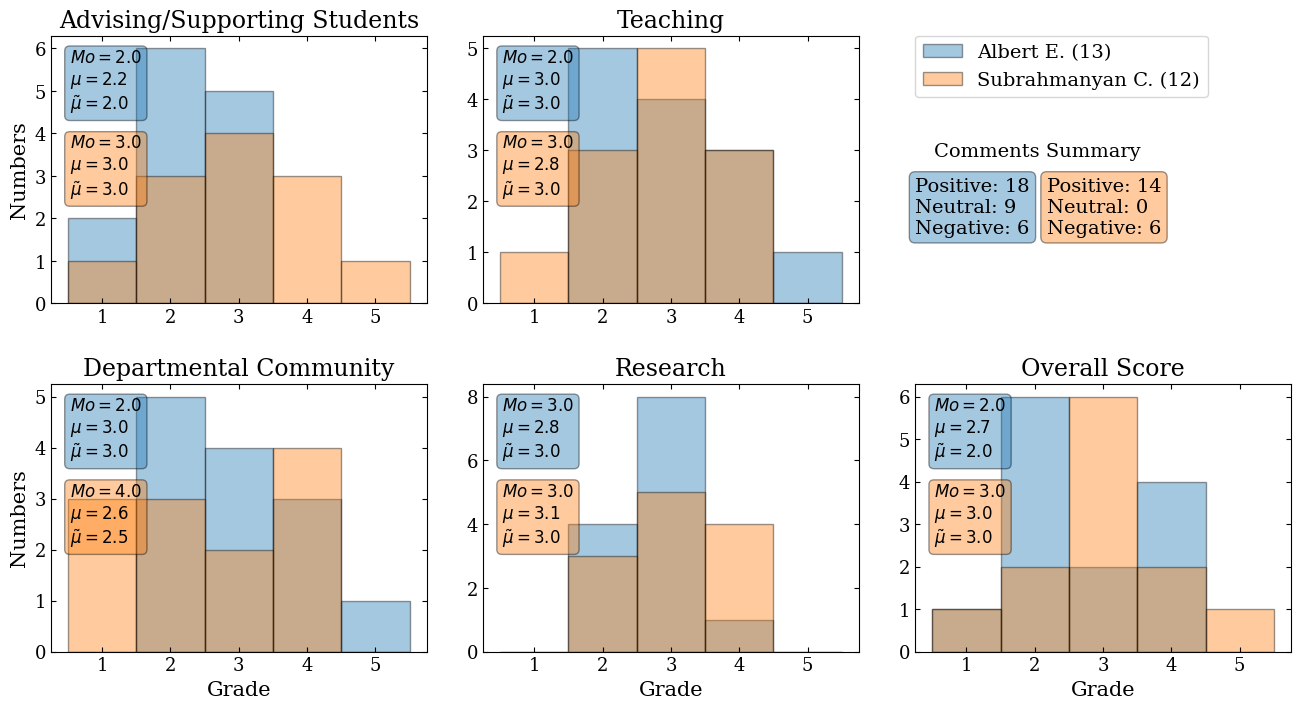}
    \caption{Histograms of grades for Albert E. and Subrahmanyan C.  The shown results are based on randomly generated mock data.}
    \label{fig:hist_2}
\end{figure*}

\begin{figure*}[htb]
    \centering
    \includegraphics[width = 0.85\textwidth]{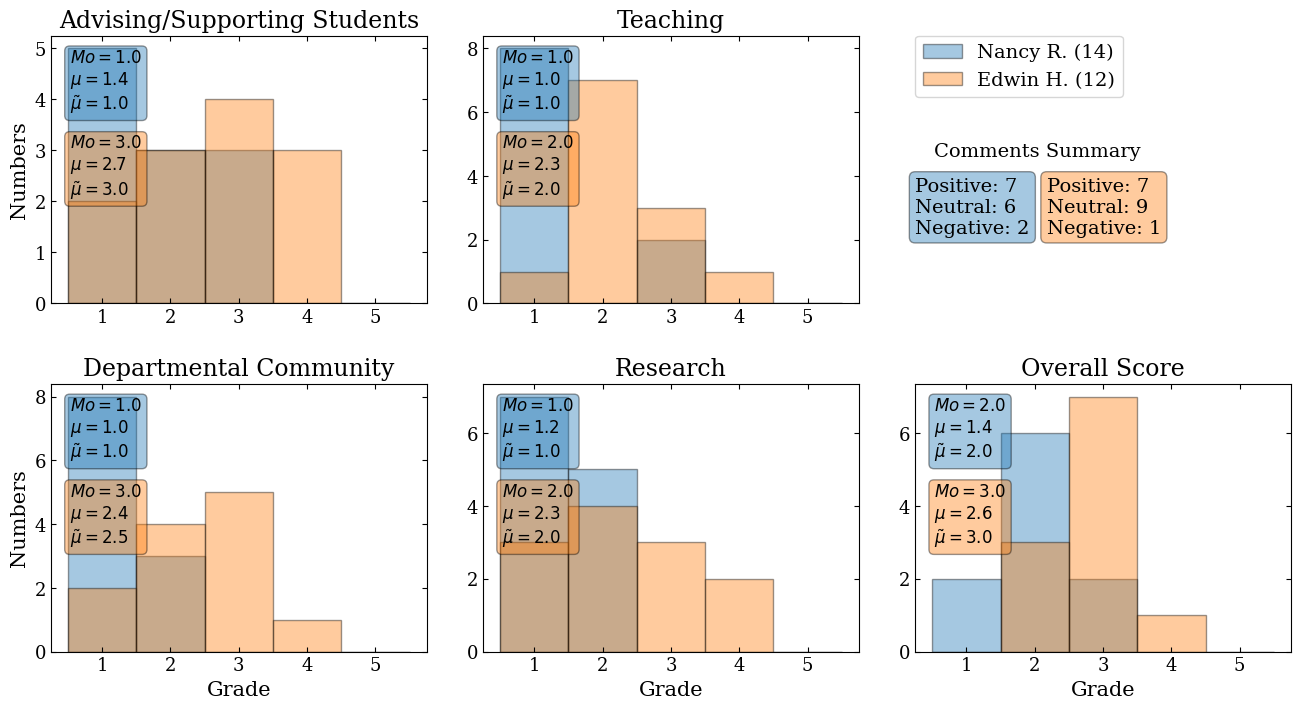}
    \caption{Histograms of grades for Nancy R. and Edwin H.  The shown results are based on randomly generated mock data.}
    \label{fig:hist_3}
\end{figure*}

In this section, we present the analysis of our survey as well as show results obtained by running our analysis pipeline on generated mock data. The real survey conducted in Yale Astronomy in spring 2022 is not included in this paper to keep its contents confidential.

The aim of this section is to demonstrate our analysis and visualization pipeline, for which we believe the mock data to suffice. We stress that results based on our mock data differ from any real survey, and dissuade from their interpretation. The generation of our mock data and associated caveats are described in Appendix \ref{sect:mockdata}.

\subsection{Numerical grades}

In Fig.~\ref{fig:mean_all_scores}, we have summarized the average values obtained for each of the broad sections on which graduate students graded the candidates. We want to note that ``Overall Score" is not the average score of the previous broad sections but was an additional question in the survey, asking students to summarize their overall opinion on whether the candidate should be hired as a faculty member (see Sect.~\ref{sect:data_collection}). The dashed lines in Fig.~\ref{fig:mean_all_scores} show the averages over all five categories for each candidate. Figure~\ref{fig:median_all_scores} depicts the same data but using the median as the measure of average instead of the arithmetic mean. We leave it to the reader to judge which measure they think is better suited for the data at hand. In order to aid in this process, we show the summarized histograms of all the candidates (split into 3-2-2) in Figs.~\ref{fig:hist_1}, \ref{fig:hist_2}, and \ref{fig:hist_3}. We did not plot more than three candidates in each panel to avoid overcrowding. 

While the mock data in Figs.~\ref{fig:hist_1}, \ref{fig:hist_2}, and \ref{fig:hist_3} more or less retrace the underlying probability distributions, real survey results can follow any distribution. In our survey, we explicitly did not curate the data from numerical evaluations (such as removing outliers) to match perceived probability distributions. 

\subsection{Individual comments}

The ASC also classified the comments received for each candidate as negative, neutral, or positive, with the goal of providing a quantitative measure intended as a quick-to-read stand-in for the general sentiment of the comments. The number of comments for each candidate in the mock data, broken down into these categories, is shown in the top right of Figs.~\ref{fig:hist_1}, \ref{fig:hist_2}, \ref{fig:hist_3}. Since this classification is inherently highly subjective, we strongly discouraged readers from relying on this subjective metric alone, and instead encouraged basing opinions on the actual comments that were included in our original report. We also considered using different metrics of concisely conveying the overall graduate student sentiment from the comments. One such idea was curating the comments for each candidate in brief summaries. However, this arguably would have increased subjectivity, and likely magnified the summaries' authors' opinions. Thus, we did not include this in our report.

Another point of consideration is that the number of comments in general does not correspond to the number of numerical grades given. In our case, roughly half of the graduate students who responded to the survey via numerical grades also added specific comments. In addition, each feedback form had a maximum of five possible text-form comment submissions. Because of this, the graduate student population that filled out the comments may significantly differ from the population that evaluated on the provided numerical scale. For example in the mock data, Grace H.'s 34 comments (see Fig.~\ref{fig:hist_1}), may be authored by anywhere between 7 to 34 graduate students (although in this example the latter would require some graduate students to only submit comments and no numerical rating, which was not possible the way our Google form was set up; see Sect.~\ref{sect:discussion}). For these reasons, while the comments certainly help convey a more detailed assessment of a candidate, they express the views of a subset of individual graduate students and thus are not necessarily representative.

\section{Survey presentation}

Once the data was analysed, the ASC compiled the results into a draft write-up. The graduate students were given the opportunity to provide feedback on the write-up and take out any comments they did not want to be included in the final report. The final report was made available to all the faculty members a few days before a meeting of all graduate students and faculty members where the ASC presented the results of the survey.

The report contained a brief description of our survey design, but focused mostly on the survey results. As such, we included the comparative plots (as presented for mock data in Sect.~\ref{sec_results}), as well as individual results for each candidates. The latter included all unedited long form comments submitted through the Google form, if not previously retracted by the respective graduate student author. Our write-up did not include any additional discussion or evaluation of the faculty candidates beyond the survey results and associated plots. 

The meeting between graduate students and faculty was centered around a presentation of the survey, its design and results, by members of the ASC. Our aim was to represent the graduate student body's assessment of the candidates as a whole. As such, we focused on the numerical evaluation visualized in the comparative plots as shown in Sect.~\ref{sec_results} for mock data. Since the individual comments were too plentiful to present in the hour-long meeting, and we wanted avoid cherry picking, we pointed the faculty to the write-up for anonymous individual opinions on the candidates. The presentation was followed by a discussion between faculty and graduate students.

\section{Discussion and recommendations for future implementations}

\label{sect:discussion}

The ASC led and conducted this survey with the primary objective of collecting student opinion on prospective faculty members in a method that maximizes fairness, transparency, and objectivity to all prospective faculty candidates as well as participating graduate students. In this section, we include our reflections on having gone through this process and our recommendations for institutions who might want to adopt this approach.


The predetermined set of questions for each candidate interview as well as the standardized survey provided graduate students the tools necessary to evaluate the different candidates consistently and objectively. Asking students to submit all questions beforehand ensured that students who might not have been comfortable speaking up at the interaction still got an opportunity to ask their questions. It also ensured that the interviews did not change drastically based on whether certain graduate students were present in a specific interview or not. Another important aspect of our discussion moderation was having rough time limits for each broad section. This ensured that we were able to cover all topics for all candidates, while still giving graduate students reasonable opportunity to ask follow-up questions. Graduate students found the time allocated for the candidate to ask questions themselves of strong use in forming an opinion.

The feedback forms were designed such that it was mandatory for respondents to assign grades for all sub-sections of the form. We feel that in future surveys, we would prefer making all grading questions optional in order to accommodate situations where respondents do not want to grade a candidate on a specific section.\footnote{In the current survey, ASC members manually deleted grades when students specifically wrote in comments that their grades for a specific section should be ignored.} In addition, the setup of the feedback form implicitly introduced equal weights to the four probed categories `Advising', `Teaching' `Community', and `Research'. Moreover, it was difficult for the evaluating students to quantify the candidate on other metrics. While the fifth question, `general assessment' was intended to provide some room for evaluation outside these four rigid categories, it was arguably limited in effect it was scored on a 0-5 scale as well, thus making this undefined grade exactly a fifth of the total score. For example, a student may strongly object a candidate's hiring for reasons that are not captured in the fixed four categories. In the most extreme case (5 points in the four categories, 1 point in `general assessment'), this would yield and average score of 4.2, hardly representative of the student's actual opinion. Possible pathways to mitigate this issue include having more categories, ideally voted on before the start of the hiring process; giving the `general assessment' score a greater weight compared to the four rigid categories; including the option to introduce custom weights to the five questions that account for the perceived importance of the corresponding categories.

While compiling the final report, the ASC focused on making its contents as data-centric as possible. We did not include any detailed analysis of the results or the comments beyond making summary plots, as shown in this paper. This was primarily because we felt including any detailed analysis or doing comment moderation/curation would include potential bias of ASC members into the process. 

Since the ASC was involved in a faculty hire for the first time, we did not have an established framework of how the survey would be done, how the results would be analyzed and presented. The ASC devised different appropriate strategies throughout the process and informed different departmental stakeholders along the way. We feel that it would have been better if we would have been able to outline the entire process in detail to the students, faculty, and candidates before the process started. 

We should also note that while graduate students presented this report to all faculty members, no students were involved in the final faculty meeting which took place to finalize the hire. Though the department had specifically reached out to students asking for feedback on the candidates, graduate students did not know beforehand how their feedback would be utilized. Having this knowledge beforehand might be beneficial in terms of both survey design and execution.

Although this survey was designed and used to collect graduate student feedback, it could also be adapted easily to collect feedback from undergraduate students, faculty and staff members. It is also important to note that this process was carried out entirely digitally. All meetings mentioned in this document happened over Zoom. Thus, there might be additional considerations that need to be made while repeating this in a non-digital setting.


\software
{Python \citep{10.5555/1593511}}, {Matplotlib \citep{Hunter2007}, Numpy \citep{harris2020array}} 

\section*{Acknowledgments}
The authors thank Michelle Nearon, Larry Gladney, Alison Coil for valuable input. The authors also emphasize the importance of all Yale Astronomy graduate students for providing feedback, as well as the Yale Astronomy faculty for inviting graduate student input.

\appendix

\section{Accessing The Survey Form \& Analysis Pipeline} \label{ap:sec:data_release}

The form used to collect survey responses as well as our analysis pipelines can be accessed at the following GitHub repository -- \url{https://github.com/aritraghsh09/asc-faculty-hire-paper-assets}

\section{Interview questions}
\label{sect:list_of_questions}

\newcommand{\edited}[1]{\textcolor{Mulberry}{\textit{#1}}} 

This appendix contains a list of questions that were asked to each faculty candidate during their meeting with the graduate students (see Sect.~\ref{ssec_questions}). For brevity, we only show the most upvoted questions, i.e. those that were always asked to each candidate. In addition, some questions were edited in order to retain confidentiality. These questions are indicated by purple coloring and italic font.

\subsection{Advising and supporting students}

\begin{enumerate}
    \item What do you think you would be able to bring to the table in terms of supporting graduate students facing non-academic challenges? (for eg. financial hardships, challenges related to family/partners, challenges related to mental health)
    \item How many graduate students (at a time) do you see yourself advising? Do you see yourself running a “group,” or treating each student and their projects independently?
    \item \edited{What specific needs do you think international students have? How would you be able to support them?
    \item In what capacity have you worked with undergraduate or graduate students?}  
    \item What is your advising style?
\end{enumerate}

\subsection{Teaching}

\begin{enumerate}
    \item If you could invent a new graduate-level class, what would the syllabus look like? 
    \item \edited{What kind of teaching pedagogy do you believe in? Are there any specific resources regarding teaching at our institution that you would be excited to use?}
\end{enumerate}

\subsection{Departmental community}

\begin{enumerate}
    \item Besides research and teaching, what aspect of the departmental community would you be interested in engaging with? Would you like to start something new? Why? What?
    \item Do you have any experience in building community? What do you think are the most crucial steps/actions needed in order to build a wholesome community?
\end{enumerate}

\subsection{Research}

\begin{enumerate}
    \item \edited{Tell us your vision for a summer project, a one year project, and a PhD thesis project?}
    \item \edited{Do you have plans to collaborate with faculty, postdocs and students from other departments at our institution? If so, how?}
    \item \edited{Given how the graduate curriculum is structured at our institution, how would you expect your students to balance their research and class/teaching responsibilities?}
\end{enumerate}

\section{Generating mock data}
\label{sect:mockdata}

Mock data was generated to approximate the results of the survey in the interest of preserving anonymity and confidentiality. We generated mock data for 7 candidates in the five categories of our original survey: Advising/Support Students, Teaching, Departmental Community, Research, and Overall Score. For each mock candidate, we selected a random number of survey responses between 11 and 15, which were the original lower and upper bounds of our responses. We generated a random average score for each candidate selected from a normal distribution with a mean of 4.0 and and standard deviation of 1.0, with a forced upper bound of 5.0. We would like to emphasize that this distribution was not selected to match the distribution of responses received during the survey. Rather, these average scores are just selected to generate mock data with a large spread for the purpose of visualization. 

Once each mock candidate is assigned an average score, we use this average to generate random data in each of the five categories for that candidate for a selected number of survey responses. For example, data for the candidate Nancy R. was generated using a normal distribution centered at a mock average of 1.69 and a standard deviation of 1.0 for 14 responses in each of the five categories. Each category has the same number of responses because each category was a required response on our survey. 

For each candidate we also generate a random number of comments classified as positive, neutral, and negative. The comments themselves are not generated, rather we simply generate a random number of comments of each type. We used the same upper and lower bounds for each type of comment as we received in the original survey. The number of positive comments ranges from 7 to 29, the number of neutral comments ranges from 0 to 12, and the number of negative comments ranges from 1 to 7. We generate 7 sets of random comments (one set per candidate), then compute an average comment score for each set. These sets are sorted by average score, and matched based on the average score of each of the seven candidates so that the higher ranked candidates based on the categorical data are allotted the higher ranked comments. This step was taken to better approximate our original data, where highly rated candidates based on the categorical scores were also highly rated based on the comments.

\bibliography{sample631}{}
\bibliographystyle{aasjournal}

\end{document}